\documentclass[apj]{emulateapj}

\begin{document}

\newcommand\rxs{1RXS J154439.4$-$112820}
\newcommand\pos{(J2000) $\alpha$=$15^{\rm h}44^{\rm m}39^{\rm s}\!.38$, $\delta$=$-11^{\circ}28^{\prime}04^{\prime \prime}\!.3$}
\newcommand\psr{PSR J1023+1038}

\submitted{Submitted to the Astrophysical Journal Letters on \today}
 
\shorttitle{NuSTAR Observations of \rxs}
\shortauthors{Bogdanov}

\title{A \textit{NuSTAR} Observation of the Gamma-Ray-Emitting X-ray Binary and \\ Transitional Millisecond Pulsar Candidate 1RXS J154439.4--112820}

\author{Slavko Bogdanov}
\affil{Columbia Astrophysics Laboratory, Columbia University, 550 West 120th Street, New York, NY 10027, USA}

\begin{abstract}  
  I present a 40 kilosecond \textit{Nuclear Spectroscopic Telescope
    Array} (\textit{NuSTAR}) observation of the recently identified
  low-luminosity X-ray binary and transitional millisecond pulsar
  (tMSP) candidate \rxs, which is associated with the high-energy
  $\gamma$-ray source 3FGL J1544.6--1125.  The system is detected up
  to $\sim$30 keV with an extension of the same power-law spectrum and
  rapid large-amplitude variability between two flux levels observed
  in soft X-rays. These findings provide further evidence that
  \rxs\ belongs to the same class of objects as the nearby bona fide
  tMSPs PSR J1023+0038 and XSS J12270--4859 and therefore almost
  certainly hosts a millisecond pulsar accreting at low luminosities.
  I also examine the long-term accretion history of \rxs\ based on
  archival optical, ultraviolet, X-ray, and $\gamma$-ray light curves
  covering the past $\sim$decade. Throughout this period, the source
  has maintained similar flux levels at all wavelengths, which is an
  indication that it has not experienced prolonged episodes of a
  non-accreting radio pulsar state but may spontaneously undergo such
  a state transformation in the future.
\end{abstract}
\keywords{pulsars: general --- stars: neutron --- X-rays: binaries}

\section{Introduction}
The recent discovery of three millisecond pulsar (MSP) binary systems
that alternate between clearly distinguishable rotation- and
accretion-powered states have revealed a new aspect of compact
binaries containing neutron stars \citep{Pap13,Pat14,Bassa14}.  In
their accreting states, these objects show rapid optical/UV
variability and a bimodal X-ray variability pattern
\citep{Lin14a,deM13,Bog15a} --- the X-ray flux switches rapidly
between two distinct high and low ``modes'' that differ by a factor of
$6-10$.  In addition, these systems constitute the only
$\gamma$-ray-emitting variety of low-mass X-ray binary; both confirmed
nearby transitional MSPs, PSR J1023+0038 and XSS J12270--4859, are
bright \textit{Fermi} LAT sources even when accreting\footnote{For the
  third transitional MSP, PSR J1824--2452I, which is situated in the
  globular cluster M28, its $\gamma$-ray emission cannot be reilabily
  disentangled from the 11 other radio MSPs in the cluster.}. The
unique combination of observational characteristics offers a fairly
straightforward way to identify additional objects that belong to this
class.

In \citet{Bog15b}, we examined the X-ray and optical/UV properties of
the bright \textit{ROSAT} source \rxs, which is positionally
coincident with the high-energy $\gamma$-ray source 3FGL J1544.6--1125
\citep{Step10,Mas13}. The variability and spectral properties of this
system were found to be essentially identical to those observed from
the nearby transitional millisecond pulsar binaries PSR J1023+0038
\citep{Bog15a} and XSS J12270--4859 \citep{deM13}.  The available
observational evidence points to \rxs~ most probably being a
transitional MSPs in a low-luminosity accreting state, making it only
the fourth such system to be identified.

Two of the transitional MSPs, XSS J12270--4859 and PSR
J1824--2452I, are currently in a dormant, accretion-disk-free state
\citep{Bassa14,Roy15,Bog14,Pap13}, leaving only PSR J1023+0038 and
\rxs\ available for further detailed studies of the accreting state.
At present, the underlying physical processes responsible for the
observed phenomenology, especially the rapid X-ray mode switching and
unexpectedly bright high-energy $\gamma$-ray emission associated with
the accreting state, are poorly understood.  In light of the
transitory nature of the accreting state, it is therefore crucially
important to investigate each system extensively at all wavelengths
before a transformation to a possibly prolonged non-accreting episode
(lasting years to decades).

On-going multiwavelength studies of PSR J1023+0038 in its present
accreting state \citep{Arch14,Del14,Bog15a} are revealing new
information regarding the accretion and jet production processes that
operate in transitional MSPs.  In this paper, I present complementary
results on \rxs\, including a target of opportunity hard X-ray
observation using the \textit{Nuclear Spectroscopic Telescope Array}
(\textit{NuSTAR}), which further confirm the similarities of this
system with transitional MSPs. I also investigate the broad band
spectral energy distribution and long-term variability of the system,
which further strenghten the classification of this system as a
transitional MSP. This work is organized as follows. In \S 2, I
summarize the \textit{NuSTAR} observation and data analysis. In \S 3,
I describe the X-ray spectroscopic analysis, while in \S 4 I
focus on the hard X-ray variability. In \S5 I discuss the broad band
spectral energy distribution of \rxs. I offer a discussion in \S6 and
conclusions in \S7.

\section{Observation and Data Analysis}
The \rxs\ system was observed with \textit{NuSTAR} \citep{Har13} on 23
March 2015 (ObsID 90001007) in a 40 ks effective on-source exposure
obtained through the Director's Discretionary Program. \textit{NuSTAR}
is the first focusing hard X-ray telescope, with sensitivity over the
3--79 keV energy range. It consists of two co-aligned telescopes,
which use grazing-incidence mirror assemblies to focus hard X-rays
onto two Focal Plane Modules, FPMA and FPMB.

The data were processed using the {\tt nupipeline} script in NuSTARDAS
and the images, spectra, and light curves were generated using the
{\tt nuproducts} task.  \rxs\ is the only hard X-ray source within the
telescope field of view and is detected with high significance up to
$\sim$30 keV.  For both the spectroscopic and time variability
analyses, the source counts were extracted from a circular region of
radius $60''$ (which encircles $\sim$80\% of the total source energy)
centered on the optical position reported in \citet{Bog15b}. The
background was taken from a $60''$ radius source-free region.


\begin{figure}
 \centering
\includegraphics[angle=270,width=0.47\textwidth]{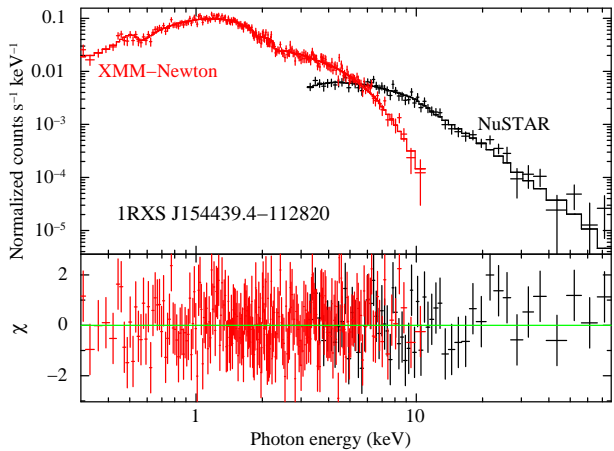}
\caption{\textit{NuSTAR} FPMA+FPMB (black) and \textit{XMM-Newton}
  MOS1+MOS2 (red) spectra of \rxs jointly fitted with an absorbed
  power-law model. The bottom panel shows the best fit residuals
  expressed in terms of $\sigma$ with error bars of size unity. See
  text for best fit parameters.}
\end{figure}

\section{X-ray Spectroscopy}
I conducted a spectral analysis in XSPEC version 12.8.2
\citep{Arnaud96}.  To maximize the photon statistics, the
FPMA and FPMB spectra and response functions were combined using the
{\tt addascaspec} command in FTOOLS.  Based on the results from the
\textit{XMM-Newton} spectroscopy presented in \citep{Bog15b}, the
\textit{NuSTAR} FPMA and FPMB spectra were fitted with an absorbed
power-law assuming the {\tt tbabs} interstellar absorption model and
chemical abundances from \citet{Wilms00}. This produces a best fit
with hydrogen column density  of $N_{\rm H}\le
1.3\times 10^{22}$ cm$^{-2}$, spectral photon index
$\Gamma=1.68\pm0.06$, an unabsorbed $3-79$ keV flux of
$F_X=(6.6\pm0.4) \times10^{-12}$ erg cm$^{-2}$ s$^{-1}$, and
$\chi^2_{\nu}=1.14$ for 156 degrees of freedom. Owing to the hard
response of \textit{NuSTAR}, the fit is insensitive to the value of
$N_{\rm}$ and only a crude upper limit is obtained. To further refine
the spectral parameters, I also conducted a joint fit to the
\textit{XMM-Newton} and \textit{NuSTAR} data.  The resulting best fit
with an absorbed power-law gives ${N_{\rm H}}=(1.7\pm0.1)\times
10^{21}$ cm$^{-2}$, $\Gamma=1.67\pm0.02$, and
$F_X=(8.4\pm0.2)\times10^{-12}$ erg cm$^{-2}$ s$^{-1}$ (0.3--79 keV),
with $\chi^2_{\nu}=0.96$ for 471 degrees of freedom.  All
uncertainties quoted are at a 90\% confidence level. No adjustment to
the cross-normalizations between the two data sets was required.  As
apparent from Figure 1, the spectral continuum shows no evidence for a
break or turn-over within the \textit{NuSTAR} band.

I also explored the possibility of spectral variability by computing
the hardness ratio variations as a function of count rate. There are
no statistically significant differences in hardness between the high
and low flux modes, which is consistent with the result from the
\textit{NuSTAR} data of PSR J1023+0038 in its accreting state
\citep{Ten14}.

\section{Hard X-ray Variability}
A total background-subtracted hard X-ray light curve was generated by
combining the emission from FPMA and FPMB.  As shown in
\citet{Bog15b}, the soft X-ray emission exhibits rapid switching
between two flux modes, with ingress and egress durations lasting of
order 10 s.  Despite the much lower count rate compared to the 0.3--10
keV \textit{XMM-Newton} light curve, the rapid variability is still
clearly apparent. The 3--79 keV light curve of \rxs, binned at 100 s
resolution, is shown in Figure 2. While the source spends most of the
time at a high flux level ($\sim$0.1 counts s$^{-1}$), multiple
instances of low flux levels (reaching down to $\sim$0.01 counts
s$^{-1}$) are evident. At the 100 s bin resolution some short-lived
low modes are likely averaged with the high modes occuring within the
same time bin. A Kuiper's test for time variability on the unbinned
time series, corrected for good time intervals by removing gaps in
exposure, gives a $9\times 10^{-16}$ ($\approx$7.9$\sigma$)
probability that a distribution with a constant count rate would
exhibit the observed level of non-uniformity.  Due to the relatively
low source count rate, the unknown orbital period, as well as known
calibration issues with the \textit{NuSTAR} on-board clock, I have not
attempted a periodicity search.

\begin{figure*}
 \centering
  \includegraphics[angle=0,width=0.75\textwidth]{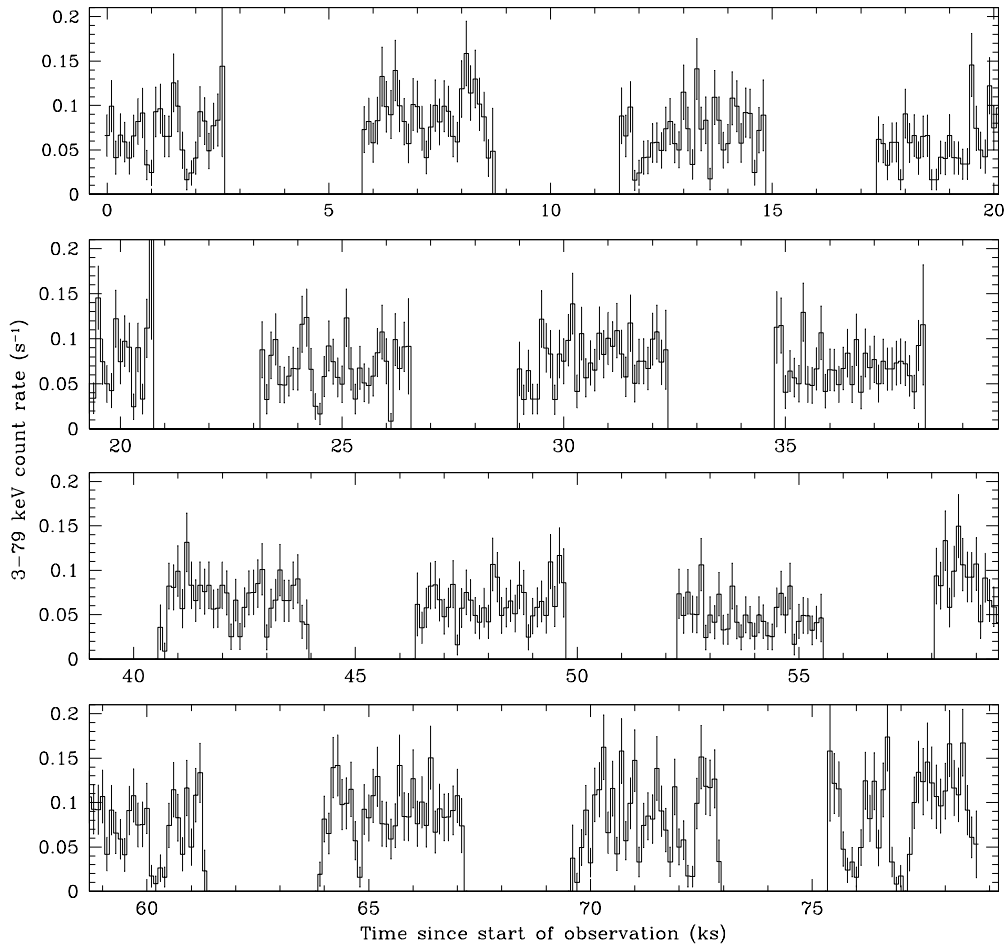}
\caption{\textit{NuSTAR} FPMA+FPMB 3--79 keV background-subtracted
  light curve of \rxs binned at a 100 s resolution. The
  regular $\sim$3 ks gaps are due to Earth occultation. Note the
  occasional short-lived rapid drops in count rate, especially towards
  the end of the observation.}
\end{figure*}

\section{Long-term Broadband Behavior}

\begin{figure}
 \centering
  \includegraphics[angle=0,width=0.47\textwidth]{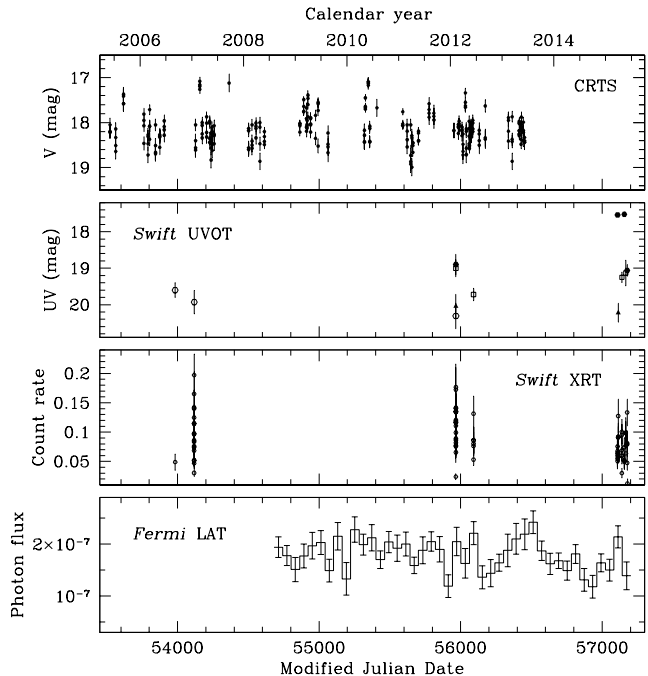}
\caption{Long-term light curves of \rxs\ from the Catalina Real-time
  Transient Survey (CRTS), \textit{Swift} UVOT, \textit{Swift} XRT
  0.3--10 keV, and \textit{Fermi} LAT 0.1--100 GeV (from top to
  bottom, respectively. The different symbols for the \textit{Swift}
  UVOT data correspond to the various filters: U (solid circles), UVW1
  (open squares), UVM2 (open circles0, and UVW2 (solid triangles).}
\end{figure}

PSR J1824--2452I in M28 was observed to be in a
very faint ($L_X\approx 10^{32}$ erg s$^{-1}$) non-accreting state in
2002 \citep{Becker03} and a low-luminosity ($L_X\approx
3\times10^{33}$ erg s$^{-1}$) accreting state in 2008
\citep{Bog11}. The optical brightness between these two states
differed by $\sim$2 magnitudes \citep{Pal13}.
As reported in \citet{Pat14}, the re-appearance of the accretion disk
in PSR J1023+0038 was accompanied by a $\sim$3.5 magnitude increase in
UV brightness and a $\sim$20-fold rise in X-ray luminosity.
Similarly, in the state transition of XSS J12270--4859
\citep{Bassa14}, the transformation from an accreting state to a radio pulsar state was signaled by a decrease in X-ray
luminosity by a factor of $\sim$10 and decline in UV and optical
brightness by $\sim$3 and $\sim$2 magnitudes, respectively.  For PSR
J1023+0038, the return of the accretion disk also coincided with a
factor of $\sim$5 enhancement in high-energy $\gamma$-ray luminosity
as seen by \textit{Fermi} LAT \citep{Stap14}. The disappearance of the
accretion disk in XSS J12270--4859 was associated with a
$\sim$2-fold decline in $\gamma$-ray flux \citep{Yi15}.

As \rxs\ closely resembles these transitional MSP systems, it may have
also experienced transitions to and from a disk-free state in the
past.  Examination of archival data is particularly valuable in this
regard, considering that both PSR J1824--2452I \citep[in
  2008-2009][]{Pal13,Lin14a} and XSS J12270--4859 \citep[in
  2012][]{Bassa14} were uncovered retroactively to have undergone
state transitions in archived data sets.  Prompted by this
possibility, I have investigated the optical/UV, X-ray, and
$\gamma$-ray variability of \rxs\ on time scales of years. For this
purpose, I use publicly available data from the Catalina Real-time
Transient Survey \citep{Drake09}, \textit{Swift} XRT and UVOT, and
\textit{Fermi} LAT data. Figure 3 shows the lightcurves from these
data sets, which are described below.

\subsection{Catalina Real-time Transient Survey}
The Catalina Real-time Transient Survey \citep[CRTS][]{Drake09}
acquires unfiltered exposures of the sky away from the Galactic plane
that usually reach equivalent $V$ filter of $\sim$19--20.  The CRTS
archive includes 324 observations of \rxs\ spanning from late 2005 to
late 2013. Removing points with magnitude errors greater than 0.2
results in 260 individual exposures.

\subsection{Swift UVOT and XRT}
The error ellipse of 3FGL J1544.6--1125 has been targeted with
\textit{Swift} once in 2006 and 2007, twice in 2012, and on six
occasions in 2015.

The \textit{Swift} UVOT data include four exposures
with the UVW1, two with the UVW1, three with the UM2, and three with
the U filter. The photometric measurements and associated
uncertainties were extracted with the {\tt uvotsource} task in FTOOLS,
using a source extraction region of radius 5$''$ centered on the
optical position given in \citet{Bog15b}.

The 12 \textit{Swift} XRT observations were all obtained in photon
counting (PC) mode. The 0.3--10 keV light curves were extracted using
the {\tt xrtpipeline} command in FTOOLS. The counts were grouped
adaptively to ensure that each time bin contained at least 20 counts.

\subsection{Fermi LAT}
To investigate the high-energy $\gamma$-ray variability of \rxs\ on long
timescales, I have retrieved all Pass 8 \textit{Fermi} Large Area
Telescope (LAT) data within $20^{\circ}$ of the source from the start
of the mission on 2008 August 4 through 2015 June 7.  The $\gamma$-ray
light curve was obtained by dividing the data set in 60 day intervals
and carrying out a binned likelihood analysis for each time segment
using the Fermi Science Tools\footnote{Available at
  \url{http://fermi.gsfc.nasa.gov/ssc/data/analysis/software/}.}
v10r0p5. Following the analysis procedures recommended by the Fermi
Science Support Center, I first generated counts, exposure, and source
maps, livetime cube, and an input source model based on the source
information provided in the 3FGL catalog \citep{Fermi15}. In the
likelihood fitting with the {\tt gtlike} task, the contribution from all
sources within $20^{\circ}$ of \rxs, as well as from the Galactic and
extragalactic diffuse emission, was taken into account. The best fit
values and uncertainties of the flux of 3FGL J1544.6--1125 were used
to construct a light curve in the 0.1--100 GeV band.

\begin{figure*}
 \centering
  \includegraphics[angle=0,width=0.8\textwidth]{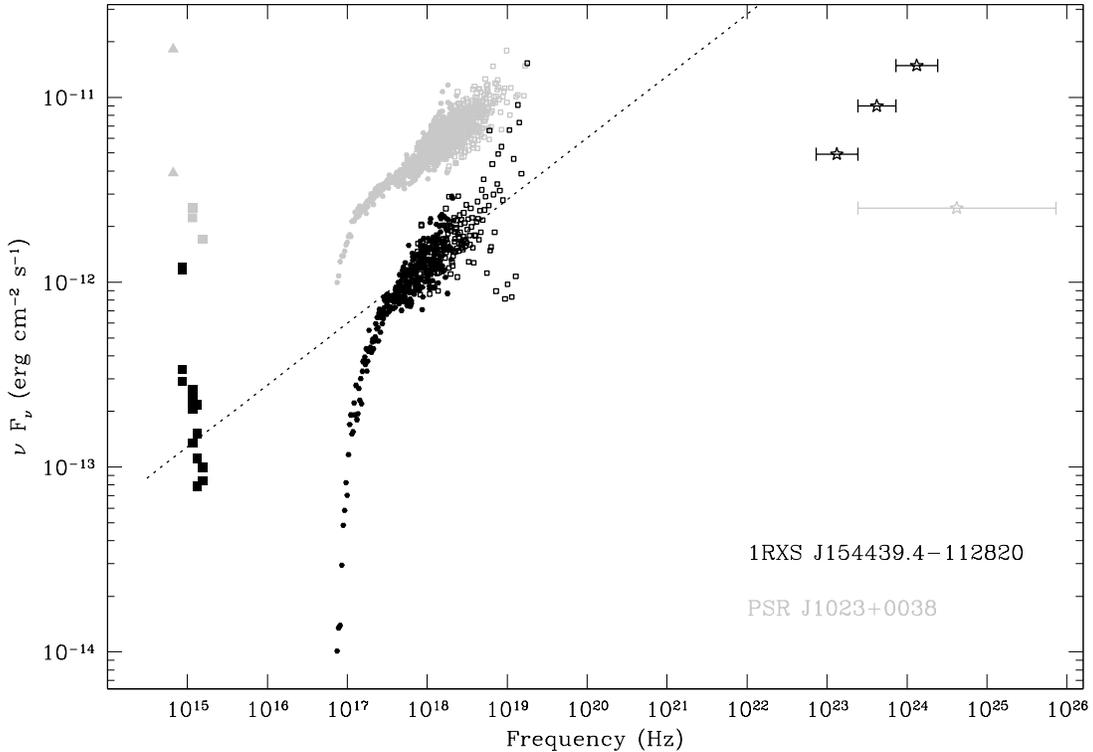}
\caption{The broadband spectrum of \rxs (black data points) covering
  the optical to 100 GeV $\gamma$-rays range, showing the
  \textit{Fermi} LAT (open stars), \textit{NuSTAR} (open squares),
  \textit{XMM-Newton} EPIC (solid circles), \textit{Swift} UVOT (solid
  squares) and \textit{XMM-Newton} OM B filter (solid triangles). The
  light grey points correspond to the spectral energy distribution of
  PSR J1023+0038 \citep[see][for details]{Bog15a}.}
\end{figure*}

Although the optical, UV and X-ray data have gaps in coverage
spanning several months to years, when combined with the $\gamma$-ray
data, they provide useful constraints on the long-term behavior of
\rxs\ within the past decade.  For instance, it is apparent that there
have been no intervals spanning months to years in which the system
has been in a low-flux, non-accreting state. The \textit{Fermi} LAT
0.1--100 GeV light curve shows no appreciable deviations from the mean
photon flux of $1.8\times10^{-7}$ cm$^{-2}$ s$^{-1}$.  This is
consistent with variability index of 47.5 reported for 3FGL
J1544.6--1125 in the \textit{Fermi} LAT 4-year catalog
\citep{Fermi15}, which indicates no statistically significant flux
changes over the course of the mission.

On the other hand, the CRTS and the most recent \textit{Swift} UVOT
UVW1 data show instances of enhancement in brightness by $\sim$$1-1.5$
magnitudes compared to the typical value.  These events may correspond
to occasional flaring episodes seen from PSR J1023+0038 \citep{Bog15a}
and XSS J12270--4859 \citep{deM13} in the optical, UV, and X-rays.
The \textit{Swift} XRT 0.3--10 keV exposures exhibit large-amplitude
variations that span a factor of $\sim$10 in count rate. The
corresponding range of fluxes are in good agreement with the high and
low mode levels observed with \textit{XMM-Newton} \citep{Bog15b}.

\section{Discussion}
The hard X-ray properties of the \rxs\ system as revealed by
\textit{NuSTAR} are virtually identical to those of the confirmed
transitional MSPs.  For example, \citet{deM10,deM13} investigated the
broad-band X-ray properties of XSS J12270--4859 during its accreting
state using data from \textit{XMM-Newton}, \textit{RXTE}, and
\textit{INTEGRAL}. The source spectrum in the 0.1--100 keV can be
fitted by a single power-law with index $\Gamma\approx 1.7$. In the
\textit{RXTE} data, XSS J12270--4859 showed the same mode switching
and flaring behavior found in soft X-rays.  \citet{Ten14} presented
\textit{NuSTAR} observations of PSR J1023+0038 in both the radio
pulsar and accreting X-ray binary states. The average hard X-ray
emission in the more luminous accreting state is well-described a
power-law with photon index $\Gamma = 1.66$ with a luminosity of
$6\times 10^{33}$ erg s$^{-1}$ (3--79 keV). Rapid variability between
two clearly separated flux levels was also found.

The available optical/UV, soft and hard X-ray and high-energy
$\gamma$-ray data for \rxs\ allows an examination of the broad-band
emission spectrum of the system. Figure 4 shows the optical to GeV
$\gamma$-ray range for \rxs. For an illustrative comparison, the
multi-wavelength data set of PSR J1023+0038 in its accreting state is
also shown. The overall similarities in the shape of the broadband
spectral energy distributions offers an additional indicator that
\rxs\ belongs to the same class of objects as PSR J1023+0038, XSS
J12270--4859, and PSR J1824--2452I. In light of this, the same
interpretations used for these objects can be invoked to explain the
different emission components of \rxs.  The optical flux is likely a
combination of emission from the companion and the accretion disk,
with the latter dominating the emission in the UV. The X-ray emission
presumably originates in the inner disk, close to or at the surface of
the neutron star.  The detection of coherent X-ray pulsations during
the high mode X-ray emission in PSR J1023+0038 \citep{Arch14} and XSS
J12270--4859 (Papitto et al.~2015) and the absence of radio pulsations
\citep{Stap14,Bog15a} suggests that at least during these intervals
active accretion onto the stellar surface is taking place and the
radio pulsar mechanism is quenched.
The cause of the aperiodic rapid drops to a low flux mode is less
clear as it may correspond to changes to a propeller ejection process
\citep{Ill75} or a dead/trapped disk regime
\citep{Sun77,Spru93,Dang10,Dang12}. Moreover, it is not clear whether
the inflow of matter is via a ``classical'' thin disk or a radiatively
inefficient accretion flow \citep{Rees82}.

The persistent $\gamma$-rays from this system possibly arise due
up-scattering of X-rays at the boundary between the accretion flow and
the neutron star magnetosphere. Two general varieties of models have
been put forth to explain the observed $\gamma$-rays, which differ
principally in the assumption of an active or inactive
rotation-powered pulsar wind.  In the active pulsar wind scenario, the
X-ray and $\gamma$-ray emission arise due to an intrabinary shock
produced at the interface of the accretion disk and the pulsar wind
\citep[][]{Tak14,Cot14}. The X-rays are
generated via synchrotron radiation in the shock, while the
$\gamma$-rays are the product of inverse Compton scattering of the
photons emanating from the accretion disk off the pulsar wind
particles.  As discussed by \citep{Pap15b}, for the case of an
inactive (``quenched'') pulsar wind, the system is assumed to be
in the so-called propeller regime in which most of the inflowing
material is expelled by the rapidly rotating pulsar magnetosphere.
The X-ray luminosity is then due to synchrotron plus disk emission,
with the flux moding arising from rapid switching between accretion
onto the star (the high mode) and propeller ejection (the low
mode). The $\gamma$-ray emission is produced by synchrotron
self-Compton emission buy accelerated particles in the turbulent
boundary between the pulsar magnetosphere and the accretion disk.

While \rxs\ appears to be a close analog of the confirmed transitional
MSPs, some notable differences are also apparent.  In particular, the
$\gamma$-ray flux of \rxs\ is $\sim$3 higher that that of PSR
J1023+0038, whereas the mean X-ray flux is $\sim$3 times lower.  While
the lower X-ray flux can be attributed to a greater distance (assuming
a comparable intrinsic luminosity), the relative differences in X-ray
to $\gamma$-ray flux may arise either due to higher efficiency of the
$\gamma$-ray production mechanism or due to a more favorable viewing
angle if the radiation pattern is not isotropic.  In addition, for PSR
J1023+0038, the high and low mode X-ray fluxes differ by a factor of
$\approx$6, while for \rxs\ the flux varies by a factor of
$\approx$10. Assuming that during the low X-ray mode, the accretion
flow is cleared out of the pulsar magnetosphere, this difference could
be ascribed to a larger accretion disk truncation radius in \rxs. If
this radius coincides with the pulsar light cylinder ($cP/2\pi$) it
implies a more slowly spinning neutron star, although other factors
like a less massive neutron star or a different magnetic field
configuration may play a role in determining the high/low flux ratio.

\section{Conclusions}
I have presented the first hard X-ray observation of the
low-luminosity X-ray binary and \textit{Fermi} LAT source, 1RXS
J154439.4--112820 / 3GL J1544.6--1125. The binary is detected at high
significance up to $\sim$30 keV with a continuation of the power-law
spectrum observed in soft X-rays.  The hard X-rays also exhibit
variability on timescales of 10--100 seconds by a factor of
$\approx$10 in flux as seen in the 0.3--10 keV range with
\textit{XMM-Newton}.  The spectral energy distribution of
\rxs\ covering the optical through  GeV range is qualitatively very
similar to that of PSR J1023+0038 \citep[as well as XSS J12270--4859;
  see][]{deM10,deM13}.  These findings  further strengthen the
argument that this system belongs to the same class of objects as PSR
J1023+0038, XSS J12270--4859 and PSR J1824--24I, namely a compact
binary hosting a MSP that is presently accreting at low luminosities.

I find no clear evidence in archival data for prolonged low flux
intervals at any wavelength implying that within the past $\sim$decade
\rxs\ has most likely not experienced a transformation to a disk-free
state.  Nevertheless, as the strongest transitional MSP candidate,
\rxs\ warrants dedicated monitoring to identify behavior that may
signal the disappearance of its accretion disk and the activation of
the rotation-powered millisecond pulsar.

\acknowledgements I thank F.~Harrison for approving the DDT request
that enabled the \textit{NuSTAR} observation.  This work is based on
data from the \textit{NuSTAR} mission, a project led by the California
Institute of Technology, managed by the Jet Propulsion Laboratory, and
funded by NASA.  I thank the \textit{NuSTAR} Operations, Software and
Calibration teams for support with the execution and analysis of these
observations.  This research has made use of the \textit{NuSTAR} Data
Analysis Software (NuSTARDAS) jointly developed by the ASI Science
Data Center (ASDC, Italy) and the California Institute of Technology
(USA), the arXiv, NASA ADS, and data and software facilities from the
Fermi Science Support Center, managed by the HEASARC at NASA GSFC.


\begin{thebibliography}{}

\bibitem[Acero et al.(2015)]{Fermi15} Acero, F., Ackermann, M., Ajello, M., et al.~2015, ApJS, 218, 23
  
\bibitem[Archibald et al.(2015)]{Arch14}  Archibald, A.~M., Bogdanov, S., Patruno, A., et al.~2015, \apj, 807, 62 

\bibitem[Arnaud(1996)]{Arnaud96} Arnaud, K. A. 1996, in ASP Conf. Ser. 101, Astronomical Data Analysis Software and Systems V, ed. G. H. Jacoby \& J. Barnes (San Francisco, CA: ASP), 17
  
\bibitem[Bassa et al.(2014)]{Bassa14} Bassa, C.~G., Patruno, A., Hessels, J.~W.~T., et al.~2014, MNRAS, 441, 1825

\bibitem[Becker et al.(2003)]{Becker03} Becker, W., Swartz, D.~A., Pavlov, G.~G., et al.~2003, \apj, 594, 798
  
\bibitem[Bogdanov et al.(2014a)]{Bog14} Bogdanov, S., Patruno, A., Archibald, A.~M., et al.~2014a, \apj, 789, 40    

\bibitem[Bogdanov et al.(2015)]{Bog15a} Bogdanov, S., Archibald, A., Bassa, C., et al.~2015, \apj, 806, 148

\bibitem[Bogdanov \& Halpern(2015)]{Bog15b} Bogdanov, S., \& Halpern, J. P.~2015, \apj, 803, L27

\bibitem[Bogdanov et al.(2011)]{Bog11} Bogdanov, S., van den Berg, M., Servillat, M., et al.~2011, 730, 81
  
\bibitem[Chakrabarty et al.(2014)]{Chak14} Chakrabarty, D., Tomsick, J.~A., Grefenstette, B.~W., et al.~2014, \apj, 797, 92
  
\bibitem[Coti Zelati et al.(2014)]{Cot14} Coti Zelati, F., Baglio, M.~C., Campana, S., et al.~2014, MNRAS, 444, 1783

\bibitem[D'Angelo \& Spruit(2010)]{Dang10} D'Angelo, C.~R., \& H. C. Spruit H.~C.~2010, MNRAS, 406, 1208

\bibitem[D'Angelo \& Spruit(2012)]{Dang12} D'Angelo, C.~R., \& H. C. Spruit H.~2012, MNRAS, 420, 416
  
\bibitem[Deller et al.(2015)]{Del14} Deller, A.~T., M\'oldon, J., Miller-Jones, J.~C.~A., et al., 2015, \apj, 809, 13

\bibitem[de Martino et al.(2013)]{deM13} de Martino, D., Belloni, T., Falanga, M., et al.~2013, A\&A, 550, 89

\bibitem[de Martino et al.(2010)]{deM10}  de Martino, D., Falanga, M., Bonnet-Bidaud, et al.~2010, A\&A, 515, 25

\bibitem[Drake et al.(2009)]{Drake09} Drake, A.~J., Djorgovski, S.~G., Mahabal, A., et al.~2009, \apj, 696, 870

\bibitem[Harrison et al.(2013)]{Har13} 	Harrison, F.~A., Craig, W.~W., Christensen, F.~E., et al.~2013, \apj, 770, 103

\bibitem[Hill et al.(2011)]{Hill11} Hill, A.~B., Szostek, A., Corbel, S., et al.~2011, MNRAS, 415, 235
  
\bibitem[Illarionov \& Sunyaev(1975)]{Ill75} Illarionov, A.~F., \& Sunyaev, R.~A.~1975, A\&A, 39, 185
  
\bibitem[Linares et al.(2014)]{Lin14a} Linares, M., Bahramian, A., Heinke, C., et al.~2014, MNRAS, 438, 251

\bibitem[Masetti et al.(2013)]{Mas13} Masetti, N., Sbarufatti, B., Parisi, P., et al.~2013, A\&A, 559, A58

\bibitem[Pallanca et al.(2013)]{Pal13} Pallanca, C., Dalessandro, E., Ferraro, F.~R., Lanzoni, B., \& Beccari, G.~2013, \apj, 773, 122

\bibitem[Papitto et al.(2013)]{Pap13} Papitto, A., Ferrigno, C., Bozzo, E., et al.~2013, Nature, 501, 517

\bibitem[Papitto et al.(2014)]{Pap14} Papitto, A., Torres, D.~F., \& Li, J.~2014, MNRAS, 438, 2105

\bibitem[Papitto et al.(2015)]{Pap15} Papitto, A., de Martino, D., Belloni, T.~M., et al.~2015, MNRAS, 449, L26

\bibitem[Papitto \& Torres(2015)]{Pap15b} Papitto, A., \& Torres, D.~F.~2015, \apj, 807, 33

  
\bibitem[Patruno et al.(2014)]{Pat14} Patruno, A., Archibald, A.~M., Hessels, J.~W.~T., et al.~2014, \apj, 781, L3

\bibitem[Rees et al.(1982)]{Rees82} Rees, M.~J., Begelman, M.~C., Blandford, R.~D., \& Phinney, E. S., 1982, Nature, 295, 17
  
\bibitem[Roy et al.(2015)]{Roy15} Roy, J., Ray, P~.S., Bhattacharyya, B., et al.~2015, \apj, 800, L12

\bibitem[Siuniaev \&Shakura(1977)]{Sun77} Siuniaev, R.~A., \& Shakura, N.~I.~1977, PAZh, 3, 262
  
\bibitem[Spruit \& Taam(1993)]{Spru93} Spruit, H.~C., \& Taam, R.~E.~1993, \apj, 402, 593
  
\bibitem[Stappers et al.(2014)]{Stap14} Stappers, B.~W., Archibald, A.~M., Hessels, J.~W.~T., et al.~2014, \apj, 790, 39

\bibitem[Stephen et al.(2010)]{Step10} Stephen, J.~B., Bassani, L., Landi, R., Malizia, A., Sguera, V., Bazzano, A., Masetti, N.~2010, MNRAS, 408, 422
  

\bibitem[Takata et al.(2014)]{Tak14} Takata, J., Li, K. L., Leung, G.~C. K., 2014, \apj, 785, 131

\bibitem[Tendulkar et al.(2014)]{Ten14} Tendulkar, S.~P., Yang, C., An, H., et al.~2014, ApJ, 791, 77


\bibitem[Yi \& Wang(2015)]{Yi15} Yi, X., \& Wang, Z.~2015, \apj, 808, 17
  
\bibitem[Wilms et al.(2000)]{Wilms00} Wilms, J., Allen, A., \& McCray, R.~2000, \apj, 542, 914
  
\end{thebibliography}
\end{document}